\documentclass[english,aps,prl,superscriptaddress,twocolumn]{revtex4-1}
\usepackage[latin9]{inputenc}
\usepackage{float}
\usepackage{amsmath}
\usepackage{amssymb}
\usepackage{graphicx}

\makeatletter
\usepackage{braket}
\usepackage{latexsym}
\usepackage{amssymb,amsmath}
\usepackage{tikz-cd}

\usepackage{ifthen}
\renewenvironment{figure}[1][]{%
 \ifthenelse{\equal{#1}{}}{%
   \@float{figure}
 }{%
   \@float{figure}[#1]%
 }%
 \centering
}{%
 \end@float
}

\makeatother

\usepackage{babel}
\begin{document}
\title{\noindent \emph{Spin fine-structure reveals bi-exciton geometry in
an organic semiconductor}}
\author{K.M. Yunusova}
\affiliation{LPS, Univ. Paris-Sud, CNRS, UMR 8502, F-91405, Orsay, France }
\author{S.L. Bayliss\textsuperscript{$^{\dagger}$}}
\affiliation{LPS, Univ. Paris-Sud, CNRS, UMR 8502, F-91405, Orsay, France }
\author{T. Chanelière}
\affiliation{Laboratoire Aimé Cotton, CNRS, Univ. Paris-Sud, ENS-Cachan, Université
Paris-Saclay, 91405, Orsay, France}
\affiliation{Univ. Grenoble Alpes, CNRS, Grenoble INP, Institut Néel, 38000 Grenoble,
France}
\author{V. Derkach}
\affiliation{O. Ya. Usikov Institute for Radiophysics and Electronics of NAS of
Ukraine 12, Acad. Proskury st., Kharkov, 61085, Ukraine }
\author{J. E. Anthony}
\affiliation{Department of Chemistry, University of Kentucky, Lexington, KY 40506-0055,
USA }
\author{A.D. Chepelianskii}
\affiliation{LPS, Univ. Paris-Sud, CNRS, UMR 8502, F-91405, Orsay, France }
\author{L. R. Weiss\textsuperscript{$^{\dagger}$}}
\affiliation{Cavendish Laboratory, J. J. Thomson Avenue, University of Cambridge,
Cambridge CB3 0HE, UK}
\begin{abstract}
In organic semiconductors, bi-excitons are key intermediates in carrier-multiplication
and exciton annihilation. Their local geometry governs their electronic
properties and yet has been challenging to determine. Here, we access
the structure of the recently discovered $S=2$ quintet bi-exciton
state in an organic semiconductor using broadband optically detected
magnetic resonance. We correlate the experimentally extracted spin
structure with the molecular crystal geometry to identify the specific
molecular pairings on which bi-exciton states reside.
\end{abstract}
\maketitle
\setlength{\abovedisplayskip}{-3pt} \setlength{\belowdisplayskip}{5pt}\setlength{\abovedisplayshortskip}{-3pt}\setlength{\belowdisplayshortskip}{5pt}
\begin{figure}[H]
\includegraphics[clip,width=0.8\columnwidth]{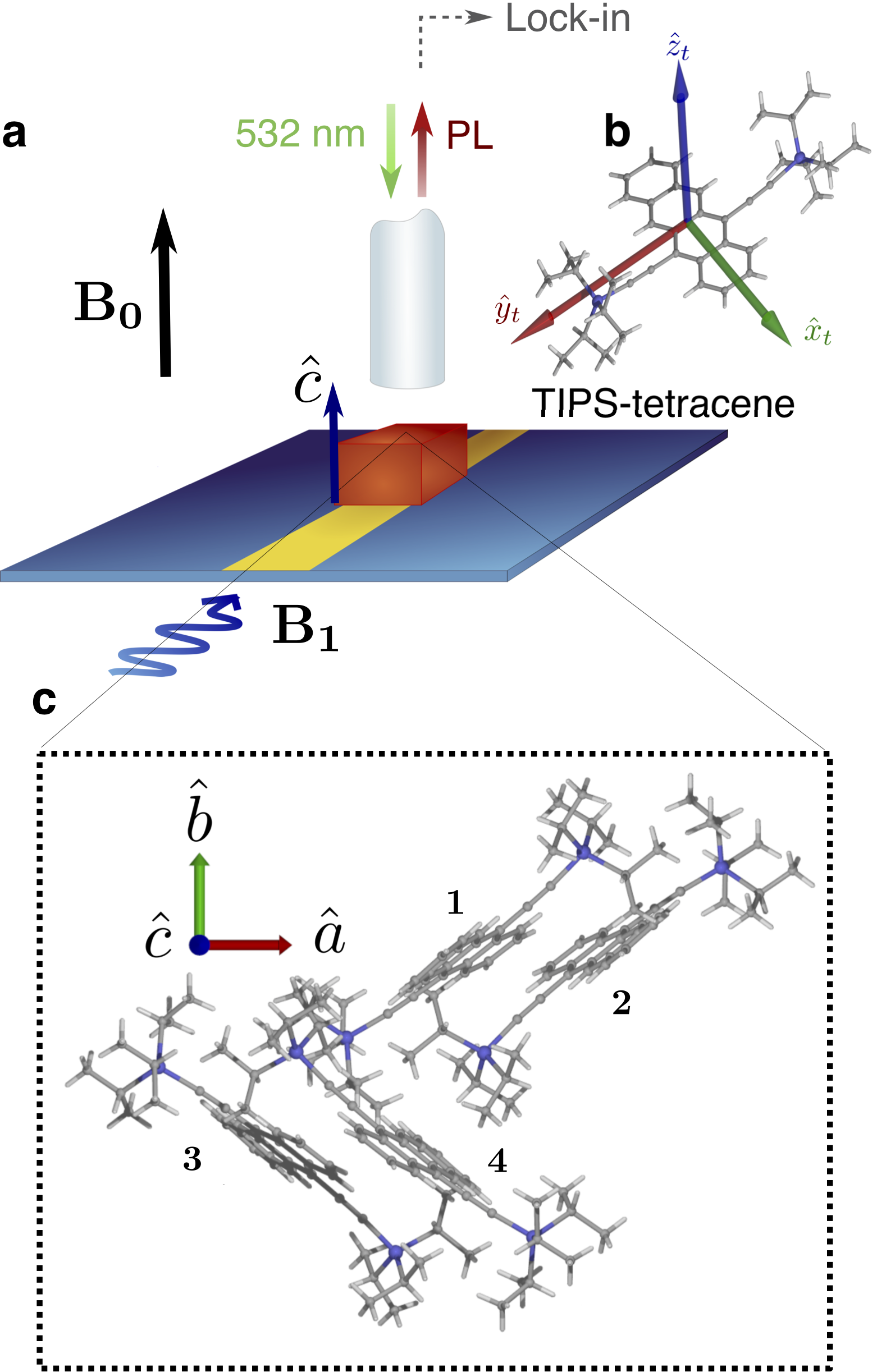}

\caption{\label{fig:fig1}Broadband ODMR of triplet-pair states.\protect \protect \\
 (a) Experimental schematic. Crystalline samples of TIPS-tetracene
(oriented with $\hat{c}$-axis as shown) were optically illuminated
under amplitude-modulated microwave excitation ($\mathbf{B}_{\mathbf{1}}$)
using a broadband strip-line in liquid helium (4 K). Photoluminescence
was collected via optical fibre to detect the microwave-induced change
in photoluminescence (PL) as a function of both microwave frequency
and static magnetic field ($\mathbf{B}_{0}$) with $\mathbf{B}_{0}\perp\mathbf{B}_{1}$.
(b) Molecular structure of TIPS-tetracene and corresponding principal
axes of the intra-triplet dipolar interaction ($\hat{x}_{t},\hat{y}_{t},\hat{z}_{t}$).
(c) Solid-state crystal structure of TIPS-tetracene with four rotationally
inequivalent molecules per unit cell labelled 1-4 and unit cell axes
($\hat{a},\hat{b},\hat{c}$). }
\end{figure}
Bi-excitons are key excited-state species in a range of nano-structured
materials from quantum-confined inorganic systems \citep{bryant1990biexciton,chen2002biexciton,hu1990biexcitons}
to synthetic molecular structures \citep{baldo2000transient,klimov1998biexcitons,masui2013analysis}.
In organic semiconductors the exciton-pair is an intermediate in both
the process of singlet fission \citep{smith2010singlet,singh1965laser,swenberg1968bimolecular}
-- the formation of a pair of spin-1 (triplet) excitons from an initial
spin-0 (singlet) excitation -- and its reverse process, triplet-triplet
annihilation \citep{singh2010photon}. While singlet fission is of
particular interest for photovoltaics \citep{rao2017harnessing,hanna2006solar,tayebjee2012thermodynamic},
where it has been shown to increase efficiencies of solar energy harvesting
beyond traditional limits \citep{einzinger2019sensitization,congreve2013external},
triplet-triplet annihilation is of interest for spectral light conversion
\citep{singh2010photon,cheng2010efficiency}, catalysis \citep{ravetz2019photoredox,khnayzer2012upconversion},
photovoltaics \citep{gray2014triplet,cheng2010efficiency}, and bio-imaging
\citep{liu2012blue,liu2018highly}.

\begin{figure*}[t]
\includegraphics[clip,width=0.95\textwidth]{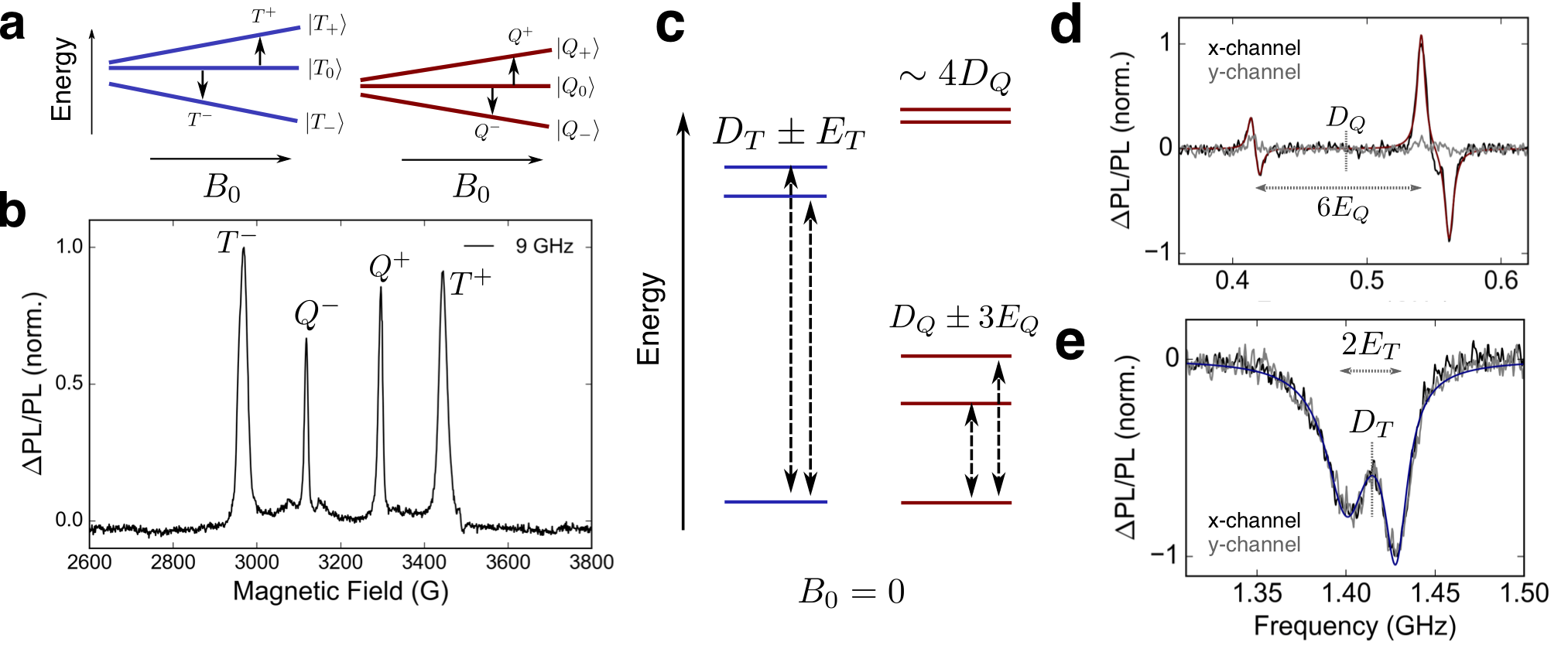}

\caption{\label{fig:fig2-new}Field-swept and zero-field ODMR of the triplet
pair state.\protect \protect \\
(a) Energy levels of the triplet and quintet $m=0,\pm1$ sub-levels
as a function of field with transitions at $9$ GHz marked with arrows
to correspond to experimentally observed transitions in (b). (b) ODMR
spectrum at $9$ GHz showing inner quintet transitions ($Q^{\pm}$)
and the outer triplet transitions ($T^{\pm}$). (c) Energy level diagram
of triplet and quintet zero-field spin sub-levels. Zero-field ODMR
spectra with quintet transitions (d) marked with corresponding
zero-field splitting parameters ($D_{Q},E_{Q}$) and simulation in
red and triplet transitions (e) marked with corresponding
triplet parameters \emph{($D_{T},E_{T}$) }and simulations in blue\emph{.}}
\end{figure*}
Despite their importance, the wavefunction of these transient, intermediate
pairs remains challenging to probe. Purely optical characterization
of bi-excitons can be ambiguous as their optical signatures typically
overlap with those of singly-excited states. Spin resonance has played
a key role in showing, unexpectedly, that in several molecular systems,
singlet fission produces a long-lived bi-exciton \citep{tayebjee2017quintet,weiss2017strongly,lubert2018multiple}.
An unambiguous signature of bi-exciton formation is the dominant exchange
interaction between the triplets within a pair (parametrized by $J\gtrsim\text{THz}$)
\citep{weiss2017strongly,bayliss2018site}. This exchange interaction
separates the pure singlet ($S=0$) from the triplet ($S=1$) and
quintet ($S=2$) pairings of the bi-exciton by $hJ$ and $3hJ$ respectively
and is identified via spin resonance or magneto-optic spectroscopy
\citep{bayliss2018site,bayliss2016spin}.

Following identification of these bi-exciton states \citep{bayliss2018site,weiss2017strongly},
we can now investigate where such bound pairs reside. Conveniently,
the $\sim\text{MHz-GHz}$ spin fine structure of the bi-exciton is
determined by inter- and intra-triplet dipolar interactions and therefore
provides a native probe of its spatial confinement and orientation
\citep{weil2007electron,benk1981theory}. We apply this approach in
TIPS-tetracene (Fig.~\ref{fig:fig1}b), a solution-processable singlet
fission material of interest for its high singlet fission efficiency
\citep{stern2017vibronically,stern2015identification}. TIPS-tetracene
is structured with side-chain modification of the canonical fission
molecule, tetracene \citep{swenberg1968bimolecular}. It crystallizes
with four orientationally inequivalent molecules per unit cell (Fig.~\ref{fig:fig1}c)
yielding six possible nearest-neighbor pair-sites within a unit cell
in addition to non-crystalline defect sites on which bi-excitons could
reside. Here we measure the spin fine structure in TIPS-tetracene
and use it to determine the molecular pair-sites where bi-excitons
reside.

The fine structure is described by the following \emph{zero-field
splitting} (ZFS) Hamiltonian

\begin{align}
\hat{H}_{zfs}/h & =\mathbf{S}^{\intercal}\cdot\mathbf{D}\cdot\mathbf{S}=D(\hat{S}_{z}^{2}-\frac{1}{3}S(S+1))+E(\hat{S}_{x}^{2}-\hat{S}_{y}^{2})\label{eq:eq1}
\end{align}
where $\mathbf{D}$ is the dipolar tensor ($D$-tensor) with parameters
$D,E$ and $\mathbf{S}$ is the relevant vector of spin operators
( with total spin $S=1,2$ for triplet, quintet states) defined along
the principal axes ($\hat{x}$, $\hat{y}$, $\hat{z}$) of the $D$-tensor.

We now show how the quintet fine-structure ($D_{Q},E_{Q}$ and the
principal axes $\hat{x}_{q},\hat{y}_{q},\hat{z}_{q}$) depends on
the underlying triplet pair orientation on two molecules (labeled
here $a$ and $b$). We assume that each triplet has the same zero-field
parameters ($D_{T}$, $E_{T}$) and differ only in orientation and
position. We define the principal axes of the first triplet state as
$(\hat{x}_{a},\hat{y}_{a},\hat{z}_{a})$ and the second triplet as
$(\hat{x}_{b},\hat{y}_{b},\hat{z}_{b})$, defined relative to the
molecular structure as in Fig.~\ref{fig:fig1}b with the vector between
them given by $\vec{r}_{ab}$ and unit vector $\hat{u}_{ab}=\text{\ensuremath{\vec{r}_{ab}/|\vec{r}_{ab}|}}$.
The zero-field Hamiltonian of the pair in the uncoupled basis is then
given by

\begin{equation}
\hat{H}_{zfs}^{(1\otimes1)}/h=\sum_{i=a,b}\mathbf{S}_{i}^{\intercal}\cdot\mathbf{D}_{T}^{i}\cdot\mathbf{S}_{i}-\Gamma(\hat{u}_{ab}\cdot\mathbf{S}_{a})(\hat{u}_{ab}\cdot\mathbf{S}_{b})+J\mathbf{S}_{a}\cdot\mathbf{S}_{b}
\end{equation}
where $\Gamma=\frac{3\mu_{0}\mu_{B}^{2}g^{2}}{4\pi|\vec{r}_{ab}|^{3}}$
gives the strength of the inter-triplet dipolar interaction with $\mu_{0}$
the magnetic permeability of free-space, $\mu_{B}$ the Bohr magneton, and
$g$ the g-factor.
In the limit of strong exchange coupling ($J\gg D_{T}$), the Hamiltonian
is approximately diagonal in the coupled spin basis defined by the
states of pure total spin \citep{bayliss2018site,bayliss2016spin,yago2016magnetic}.
Converting the above Hamiltonian to the coupled basis and projecting
into the $S=2$ subspace gives the quintet zero-field fine-structure
Hamiltonian as

\begin{equation}
\hat{H}_{zfs}^{(2)}/h=\mathbf{S}^{\intercal}\cdot\mathbf{D}_{Q}\cdot\mathbf{S}
\end{equation}
where $\mathbf{S}=(\text{\ensuremath{\hat{S}_{x},\hat{S}_{y},\hat{S}_{z}}})$
are the Pauli spin operators for total spin-2. The quintet zero-field
tensor $\mathbf{D}_{Q}$ in terms of the underlying triplet fine structure,
inter-triplet distance and dipolar interaction is given by

\begin{align}
\mathbf{D}_{Q} & =\frac{D_{T}}{6}(\sum_{i=a,b}\hat{z}_{i}\hat{z}_{i}^{\intercal}-\frac{2}{3}\hat{I}_{3})\label{eq:quintet-ham-1}\\
+ & \frac{E_{T}}{6}\sum_{i=a,b}(\hat{x}_{i}\hat{x}_{i}^{\intercal}-\hat{y}_{i}\hat{y}_{i}^{\intercal})-\frac{\Gamma}{3}(\hat{u}_{ab}\hat{u}_{ab}^{T}-\frac{1}{3}\hat{I}_{3})\nonumber
\end{align}
where $\hat{I}_{3}$ is the identify matrix in three dimensions (a
detailed derivation is in the Supplemental Material \citep{SM}).
Converting $\hat{H}_{zfs}^{(2)}$ to the form of Eq. \ref{eq:eq1}
yields the quintet dipolar parameters $D_{Q}$, $E_{Q}$, and the
principal axes $\hat{x}_{q},\hat{y}_{q},\hat{z}_{q}$ (the eigenvectors
of $\mathbf{D}_{Q}$).

\begin{figure}[!t]
\includegraphics[clip,width=1\columnwidth]{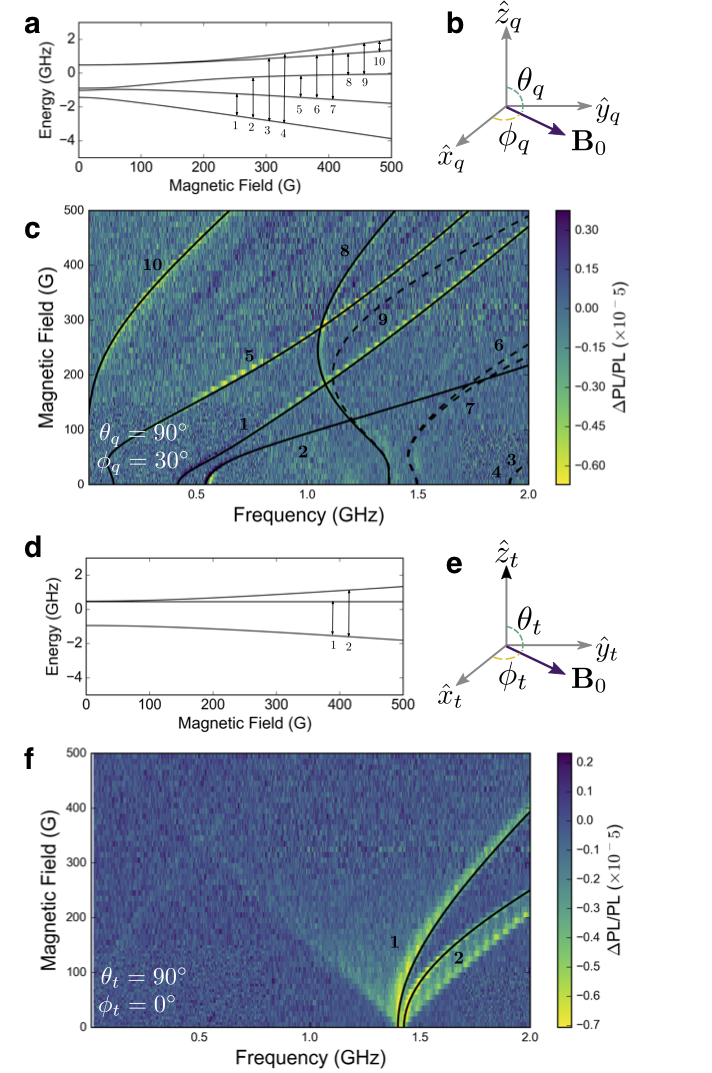}

\caption{\label{fig:fig2}Fine-structure tensors from broadband ODMR.\protect
\protect \\
 (a,d) Energy level diagram for the quintet (a) and triplet (d) states
as a function of magnetic field. Arrows indicate potential transitions,
corresponding to lines in (c) and (f) respectively. (b,e) Schematic
representation of the orientation of $\mathbf{B}$ in the quintet
(b) and triplet (e) fine-structure axes. (c) ODMR transitions associated
with the quintet state with overlay of simulated transitions. Signal
has been isolated by subtracting a scaled out-of-phase (Y)-channel
signal from the in-phase signal to remove triplet contributions. Black
lines show simulations given for $\theta_{q}=90^{\circ}$ and $\phi_{q}=30^{\circ}$
with an uncertainty of $\pm5^{\circ}$ where solid (dashed) lines
overlay (un-)observed transitions. (f) Y-channel (out-of-phase) ODMR
map of the triplet state with overlay of calculated transitions in
black with $\theta_{t}=90^{\circ}$ and $\phi_{t}=0$. }
\end{figure}
Each distinct potential pair site can be identified by its unique
fine structure parameters in a single crystal with the relation given
by Eq. \ref{eq:quintet-ham-1}. With this motivation we use a macroscopic
crystal ($\sim$mm-scale single-crystalline domain) and measure the
principal values and axes of the $D$-tensors of the triplet and quintet
states using broadband optically detected magnetic resonance (ODMR).
The experimental setup is shown in Fig.~\ref{fig:fig1}a and includes
532~nm continuous-wave (CW) light excitation, static magnetic field
($\mathbf{B}_{0}$) , and microwave radiation ($\mathbf{B}_{1}$)
with variable frequency delivered through a broadband copper strip-line.
The TIPS-tetracene crystal is aligned with $\mathbf{B}_{0}\parallel\hat{c}$.
The ODMR signal is measured by lock-in detection of microwave-induced
changes in photoluminescence (PL).

We first perform fixed-frequency ($9$ GHz), field-swept ODMR. In
agreement with previous measurements using transient electron spin
resonance, we observe two pairs of spin-transitions consistent with
the $\Delta m=\pm1$ transitions of the $S=1$ triplet exciton (which
we label $T^{\pm}$) and the $\Delta m=\pm1$ transitions of the $S=2$
quintet state (labeled $Q^{\pm}$), as shown in the Fig.~\ref{fig:fig2-new}a,b.
(Note that this spectrum confirms the expected orientation of the
crystal aligned with $\mathbf{B}_{0}\parallel\hat{c}$ as noted in
the Supplemental Material \citep{SM}) We correlate these observed
high-field transitions with zero-field transitions (magnetic field
strength $\mathbf{B}_{0}=0$, Fig.~\ref{fig:fig2-new}c,d), measured here
to sensitively determine the zero-field splitting parameters. We now
describe the microwave transitions observed experimentally. Two of
the triplet energy levels are separated by $|hD_{T}|$ from the lowest
level, and the two upper eigenstates are further split by $|2E_{T}h|$
(Fig.~\ref{fig:fig2-new}c in blue). ODMR then occurs at microwave
frequencies $\nu=D_{T}\pm E_{T}$. The three lowest quintet levels
are split by $|D_{Q}|$ from the ground state to the first two states
with a further splitting of $|6E_{Q}|$ between those two upper levels
(Fig.~\ref{fig:fig2-new}c in red). This leads to ODMR transition
frequencies at $\nu=D_{Q}\pm3E_{Q}$. Note that the previously reported
D-parameters for TIPS-tetracene are $D_{T}\sim1.4$ GHz and $D_{Q}\sim D_{T}/3$
\citep{weiss2017strongly,bayliss2014geminate}.

The spectra of triplets and bi-excitons can be separated in ODMR using
the difference in lifetime of the two species \citep{weiss2017strongly,bayliss2014geminate}.
The microwave amplitude modulation frequency ($137\;{\rm Hz}$) is
chosen such that the triplet signal appears with equal amplitude on
the in-phase (X-channel) and out-of-phase (Y-channel) lock-in channels,
which corresponds to the inverse lifetime of the triplets. The signal
from shorter lived bi-excitons appears only on the X-channel and can
be isolated by subtracting X and Y channels. The zero-field X- and
Y-channel ODMR spectra are plotted in Fig.~\ref{fig:fig2-new}d,e
in black (X-channel) and grey (Y-channel). The transitions on the
Y-channel are consistent with triplets with $|D_{T}|=1.4$ GHz and
$|E_{T}|=14$ MHz (Fig.~\ref{fig:fig2-new}e,with overlaid spectral
fit in blue). Transitions in the frequency region expected for the
quintet only appear on the X-channel and give $|D_{Q}|=477$ MHz and
$|E_{Q}|=22$ (Fig.~\ref{fig:fig2-new}d, with overlaid spectral
fit in red). The measurement of the $E$-parameters here is made possible
by the reduced linewidths observed at zero-field relative to previous
measurements under non-zero magnetic field. (Note that the spectral
fit in Fig.~\ref{fig:fig2-new}d includes a minor species with slightly
larger quintet parameters ($|D_{Q}|=490,|E_{Q}|=24$) that quickly
decays in intensity with field.)

Having extracted the principal components of the triplet and quintet
fine-structure at zero-field, we now map the resonance frequencies
as a function of magnetic field to determine the corresponding orientations
of the principal axes. The experimental ODMR maps for quintet and
triplet states are shown in Fig.~\ref{fig:fig2}c,f. The observed
resonances cannot be fit by a spin-1 state, which further confirms
the assignment (see Supplemental Material \citep{SM}). We parametrize
the orientation of the principal axes relative to the magnetic field
with the polar angle $\theta$ and azimuthal angle $\phi$ as shown
in Fig.~\ref{fig:fig2}b,e. The orientation of the quintet fine-structure
axes is obtained by fitting these maps with the spin transitions predicted
by the fine structure parameters determined at zero-field with the
orientation as input. There are 10 possible transitions between the
five quintet spin sub-levels (Fig.~\ref{fig:fig2}a), which are overlaid
on the quintet ODMR map (Fig.~\ref{fig:fig2}c). It should be noted
that the visibility of transitions depends on populations and selection
rules, and transitions 3,4,6,7 and 9 are not clearly observed experimentally.
The quintet state is oriented with fixed $\theta_{q}=90\pm5^{\circ}$
between $\hat{z}_{q}$ and $\mathbf{B}_{0}$ and $\phi_{q}=30\pm5^{\circ}$
between $\hat{x}_{q}$ and $\mathbf{B}_{0}$.

The evolution of the triplet zero-field transitions ($\sim1.4$ GHz)
with field, shown in Fig.~\ref{fig:fig2}f are consistent with $\theta_{t}=90^{\circ}$
(simulated transitions shown in black). We also observe $\theta_{t}\sim0$
peaks due to a weak powder background, which decays quickly with field.
The dominant $\theta_{t}=90^{\circ}$ triplet orientation correlates
with the high-field spectrum (Fig.~\ref{fig:fig2-new}b): triplet
peaks are separated in field by $\sim hD/g\mu_{B}$, which occurs
when $\theta_{t}\sim90^{\circ}$, whereas no peaks are observed for
$\theta_{t}\sim0$ (separation in field of $\sim2D/g\mu_{B}$). Note
that the transitions are consistent with $\phi_{t}\sim0$, but this
angle could not be extracted reliably and is not required for subsequent
analysis because the triplet states are nearly axially symmetric (i.e.,
$E_{T}\approx0$). As the D-tensor principal values and axes in the
laboratory frame are obtained from a crystalline sample aligned with
$\mathbf{B}_{0}\parallel\hat{c}$, we can now compare them with the
theoretically predicted $D$-tensors in the TIPS-tetracene crystal
structure.

\begin{figure}[!t]
\includegraphics[clip,width=1\columnwidth]{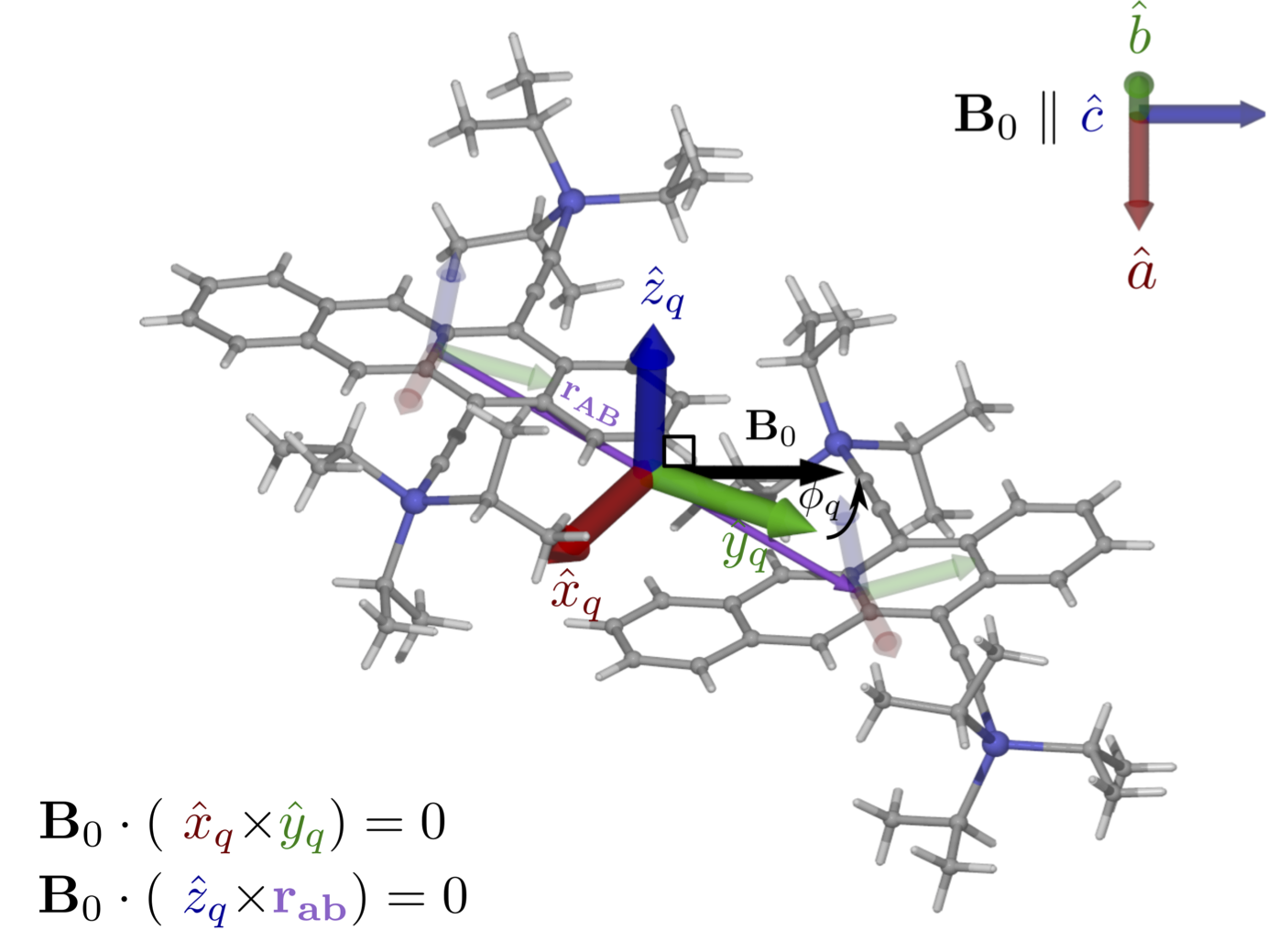}

\caption{\label{fig:fig4}Local geometry of the quintet fine-structure tensor
in the TIPS-tetracene crystal with an angle of $\theta_{q}=91.6^{\circ}$
and $\phi_{q}=30.6^{\circ}$, where the crystal is oriented with $\mathbf{B}_{0}\parallel\hat{c}$
with two co-planar triads of vectors, $(\hat{x}_{q},\hat{y}_{q},\mathbf{B}_{0})$
and $(\hat{z}_{q},\mathbf{r}_{ab},\mathbf{B}_{0})$.}
\end{figure}
There are six potential nearest-neighbor dimer configurations in the
TIPS-tetracene crystal structure (see Fig.~\ref{fig:fig1}c) and
for each we can calculate the fine structure parameters ($D_{Q},E_{Q},\theta_{q},\phi_{q}$)
using Eq.~(\ref{eq:quintet-ham-1}) and the point-dipole approximation,
the dipolar axes shown in Fig.~\ref{fig:fig1}b and the intermolecular
distances extracted from the crystal structure \citep{anthony2013Experimental}.
The full set of values are summarized in the Supplemental Material
\citep{SM}. The observed quintet parameters and extracted
angles of $\theta_{q}\sim90^{\circ}$ and $\phi_{q}\sim30^{\circ}$
are consistent with exchange-coupled triplets localized on dimers
with molecules 1,2 and 3,4 (labelled in Fig.~\ref{fig:fig1}c). The
extracted local quintet fine structure is visualized in Fig.~\ref{fig:fig4}
where the full quintet and triplet dipolar interactions are shown
with respect to the magnetic field in the lab frame and crystallographic
axes, summarizing the local structure of quintet and triplet states
in TIPS-tetracene and their relation to intermolecular geometry.

We have shown how the sensitivity of broadband magnetic resonance
enables identification of triplet-pair geometries in a material with
many possible intermolecular configurations. This approach is broadly
applicable even in cases where the crystal structure is not known
or relevant (as in disordered materials and devices with trap sites).
Here we find that the triplet pairs are localized on the closest $\pi$-stacked
dimers of the crystal structure. As the fine structure is consistent
with minimal modification from the ground-state crystal structure,
these results suggest that geometric reorganization is negligible
in the quintet bi-exciton excited-state. This description of the geometry
of the triplet pair sets the foundation for time-resolved measurements
to allow investigation of the transient localization and molecular
reorganization of the pair-state. While here we have examined a singlet
fission material, this technique is broadly applicable to any device
architecture with interacting excited states, from those incorporating two-dimensional materials and quantum dots to emerging hybrid organic-inorganic and bio-engineered structures.

\noindent \emph{Acknowledgements} We are thankful for insightful discussions
with H. Bouchiat and R.H. Friend and acknowledge support from ANR
SPINEX and Labex ANR-10-LABX-0039-PALM. L.R. Weiss acknowledges support
from Clare College, Cambridge. The authors declare no competing financial
interests. Correspondence should be addressed to A.D.C. (chepelianskii@lps.u-psud.fr)
and L.R.W (lrweiss@uchicago.edu).

\noindent $^{\dagger}$Present address: Pritzker School of Molecular
Engineering, University of Chicago, Chicago, Illinois 60637, USA.

\bibliographystyle{apsrev4-1}
\bibliography{references}

\begin{thebibliography}{34}%
\makeatletter
\providecommand \@ifxundefined [1]{%
 \@ifx{#1\undefined}
}%
\providecommand \@ifnum [1]{%
 \ifnum #1\expandafter \@firstoftwo
 \else \expandafter \@secondoftwo
 \fi
}%
\providecommand \@ifx [1]{%
 \ifx #1\expandafter \@firstoftwo
 \else \expandafter \@secondoftwo
 \fi
}%
\providecommand \natexlab [1]{#1}%
\providecommand \enquote  [1]{``#1''}%
\providecommand \bibnamefont  [1]{#1}%
\providecommand \bibfnamefont [1]{#1}%
\providecommand \citenamefont [1]{#1}%
\providecommand \href@noop [0]{\@secondoftwo}%
\providecommand \href [0]{\begingroup \@sanitize@url \@href}%
\providecommand \@href[1]{\@@startlink{#1}\@@href}%
\providecommand \@@href[1]{\endgroup#1\@@endlink}%
\providecommand \@sanitize@url [0]{\catcode `\\12\catcode `\$12\catcode
  `\&12\catcode `\#12\catcode `\^12\catcode `\_12\catcode `\%12\relax}%
\providecommand \@@startlink[1]{}%
\providecommand \@@endlink[0]{}%
\providecommand \url  [0]{\begingroup\@sanitize@url \@url }%
\providecommand \@url [1]{\endgroup\@href {#1}{\urlprefix }}%
\providecommand \urlprefix  [0]{URL }%
\providecommand \Eprint [0]{\href }%
\providecommand \doibase [0]{http://dx.doi.org/}%
\providecommand \selectlanguage [0]{\@gobble}%
\providecommand \bibinfo  [0]{\@secondoftwo}%
\providecommand \bibfield  [0]{\@secondoftwo}%
\providecommand \translation [1]{[#1]}%
\providecommand \BibitemOpen [0]{}%
\providecommand \bibitemStop [0]{}%
\providecommand \bibitemNoStop [0]{.\EOS\space}%
\providecommand \EOS [0]{\spacefactor3000\relax}%
\providecommand \BibitemShut  [1]{\csname bibitem#1\endcsname}%
\let\auto@bib@innerbib\@empty
\bibitem [{\citenamefont {Bryant}(1990)}]{bryant1990biexciton}%
  \BibitemOpen
  \bibfield  {author} {\bibinfo {author} {\bibfnamefont {G.~W.}\ \bibnamefont
  {Bryant}},\ }\href@noop {} {\bibfield  {journal} {\bibinfo  {journal} {Phys.
  Rev. B}\ }\textbf {\bibinfo {volume} {41}},\ \bibinfo {pages} {1243}
  (\bibinfo {year} {1990})}\BibitemShut {NoStop}%
\bibitem [{\citenamefont {Chen}\ \emph {et~al.}(2002)\citenamefont {Chen},
  \citenamefont {Stievater}, \citenamefont {Batteh}, \citenamefont {Li},
  \citenamefont {Steel}, \citenamefont {Gammon}, \citenamefont {Katzer},
  \citenamefont {Park},\ and\ \citenamefont {Sham}}]{chen2002biexciton}%
  \BibitemOpen
  \bibfield  {author} {\bibinfo {author} {\bibfnamefont {G.}~\bibnamefont
  {Chen}}, \bibinfo {author} {\bibfnamefont {T.}~\bibnamefont {Stievater}},
  \bibinfo {author} {\bibfnamefont {E.}~\bibnamefont {Batteh}}, \bibinfo
  {author} {\bibfnamefont {X.}~\bibnamefont {Li}}, \bibinfo {author}
  {\bibfnamefont {D.}~\bibnamefont {Steel}}, \bibinfo {author} {\bibfnamefont
  {D.}~\bibnamefont {Gammon}}, \bibinfo {author} {\bibfnamefont
  {D.}~\bibnamefont {Katzer}}, \bibinfo {author} {\bibfnamefont
  {D.}~\bibnamefont {Park}}, \ and\ \bibinfo {author} {\bibfnamefont
  {L.}~\bibnamefont {Sham}},\ }\href@noop {} {\bibfield  {journal} {\bibinfo
  {journal} {Phys. Rev. Lett.}\ }\textbf {\bibinfo {volume} {88}},\ \bibinfo
  {pages} {117901} (\bibinfo {year} {2002})}\BibitemShut {NoStop}%
\bibitem [{\citenamefont {Hu}\ \emph {et~al.}(1990)\citenamefont {Hu},
  \citenamefont {Koch}, \citenamefont {Lindberg}, \citenamefont
  {Peyghambarian}, \citenamefont {Pollock},\ and\ \citenamefont
  {Abraham}}]{hu1990biexcitons}%
  \BibitemOpen
  \bibfield  {author} {\bibinfo {author} {\bibfnamefont {Y.}~\bibnamefont
  {Hu}}, \bibinfo {author} {\bibfnamefont {S.}~\bibnamefont {Koch}}, \bibinfo
  {author} {\bibfnamefont {M.}~\bibnamefont {Lindberg}}, \bibinfo {author}
  {\bibfnamefont {N.}~\bibnamefont {Peyghambarian}}, \bibinfo {author}
  {\bibfnamefont {E.}~\bibnamefont {Pollock}}, \ and\ \bibinfo {author}
  {\bibfnamefont {F.~F.}\ \bibnamefont {Abraham}},\ }\href@noop {} {\bibfield
  {journal} {\bibinfo  {journal} {Phys. Rev. Lett.}\ }\textbf {\bibinfo
  {volume} {64}},\ \bibinfo {pages} {1805} (\bibinfo {year}
  {1990})}\BibitemShut {NoStop}%
\bibitem [{\citenamefont {Baldo}\ \emph {et~al.}(2000)\citenamefont {Baldo},
  \citenamefont {Adachi},\ and\ \citenamefont {Forrest}}]{baldo2000transient}%
  \BibitemOpen
  \bibfield  {author} {\bibinfo {author} {\bibfnamefont {M.~A.}\ \bibnamefont
  {Baldo}}, \bibinfo {author} {\bibfnamefont {C.}~\bibnamefont {Adachi}}, \
  and\ \bibinfo {author} {\bibfnamefont {S.~R.}\ \bibnamefont {Forrest}},\
  }\href@noop {} {\bibfield  {journal} {\bibinfo  {journal} {Phys. Rev. B}\
  }\textbf {\bibinfo {volume} {62}},\ \bibinfo {pages} {10967} (\bibinfo {year}
  {2000})}\BibitemShut {NoStop}%
\bibitem [{\citenamefont {Klimov}\ \emph {et~al.}(1998)\citenamefont {Klimov},
  \citenamefont {McBranch}, \citenamefont {Barashkov},\ and\ \citenamefont
  {Ferraris}}]{klimov1998biexcitons}%
  \BibitemOpen
  \bibfield  {author} {\bibinfo {author} {\bibfnamefont {V.}~\bibnamefont
  {Klimov}}, \bibinfo {author} {\bibfnamefont {D.}~\bibnamefont {McBranch}},
  \bibinfo {author} {\bibfnamefont {N.}~\bibnamefont {Barashkov}}, \ and\
  \bibinfo {author} {\bibfnamefont {J.}~\bibnamefont {Ferraris}},\ }\href@noop
  {} {\bibfield  {journal} {\bibinfo  {journal} {Phys. Rev. B}\ }\textbf
  {\bibinfo {volume} {58}},\ \bibinfo {pages} {7654} (\bibinfo {year}
  {1998})}\BibitemShut {NoStop}%
\bibitem [{\citenamefont {Masui}\ \emph {et~al.}(2013)\citenamefont {Masui},
  \citenamefont {Nakanotani},\ and\ \citenamefont
  {Adachi}}]{masui2013analysis}%
  \BibitemOpen
  \bibfield  {author} {\bibinfo {author} {\bibfnamefont {K.}~\bibnamefont
  {Masui}}, \bibinfo {author} {\bibfnamefont {H.}~\bibnamefont {Nakanotani}}, \
  and\ \bibinfo {author} {\bibfnamefont {C.}~\bibnamefont {Adachi}},\
  }\href@noop {} {\bibfield  {journal} {\bibinfo  {journal} {Org. Electron.}\
  }\textbf {\bibinfo {volume} {14}},\ \bibinfo {pages} {2721} (\bibinfo {year}
  {2013})}\BibitemShut {NoStop}%
\bibitem [{\citenamefont {Smith}\ and\ \citenamefont
  {Michl}(2010)}]{smith2010singlet}%
  \BibitemOpen
  \bibfield  {author} {\bibinfo {author} {\bibfnamefont {M.~B.}\ \bibnamefont
  {Smith}}\ and\ \bibinfo {author} {\bibfnamefont {J.}~\bibnamefont {Michl}},\
  }\href@noop {} {\bibfield  {journal} {\bibinfo  {journal} {Chem. Rev.}\
  }\textbf {\bibinfo {volume} {110}},\ \bibinfo {pages} {6891} (\bibinfo {year}
  {2010})}\BibitemShut {NoStop}%
\bibitem [{\citenamefont {Singh}\ \emph {et~al.}(1965)\citenamefont {Singh},
  \citenamefont {Jones}, \citenamefont {Siebrand}, \citenamefont {Stoicheff},\
  and\ \citenamefont {Schneider}}]{singh1965laser}%
  \BibitemOpen
  \bibfield  {author} {\bibinfo {author} {\bibfnamefont {S.}~\bibnamefont
  {Singh}}, \bibinfo {author} {\bibfnamefont {W.}~\bibnamefont {Jones}},
  \bibinfo {author} {\bibfnamefont {W.}~\bibnamefont {Siebrand}}, \bibinfo
  {author} {\bibfnamefont {B.}~\bibnamefont {Stoicheff}}, \ and\ \bibinfo
  {author} {\bibfnamefont {W.}~\bibnamefont {Schneider}},\ }\href@noop {}
  {\bibfield  {journal} {\bibinfo  {journal} {J. Chem. Phys.}\ }\textbf
  {\bibinfo {volume} {42}},\ \bibinfo {pages} {330} (\bibinfo {year}
  {1965})}\BibitemShut {NoStop}%
\bibitem [{\citenamefont {Swenberg}\ and\ \citenamefont
  {Stacy}(1968)}]{swenberg1968bimolecular}%
  \BibitemOpen
  \bibfield  {author} {\bibinfo {author} {\bibfnamefont {C.}~\bibnamefont
  {Swenberg}}\ and\ \bibinfo {author} {\bibfnamefont {W.}~\bibnamefont
  {Stacy}},\ }\href@noop {} {\bibfield  {journal} {\bibinfo  {journal} {Chem.
  Phys. Lett.}\ }\textbf {\bibinfo {volume} {2}},\ \bibinfo {pages} {327}
  (\bibinfo {year} {1968})}\BibitemShut {NoStop}%
\bibitem [{\citenamefont {Singh-Rachford}\ and\ \citenamefont
  {Castellano}(2010)}]{singh2010photon}%
  \BibitemOpen
  \bibfield  {author} {\bibinfo {author} {\bibfnamefont {T.~N.}\ \bibnamefont
  {Singh-Rachford}}\ and\ \bibinfo {author} {\bibfnamefont {F.~N.}\
  \bibnamefont {Castellano}},\ }\href@noop {} {\bibfield  {journal} {\bibinfo
  {journal} {Coord. Chem. Rev.}\ }\textbf {\bibinfo {volume} {254}},\ \bibinfo
  {pages} {2560} (\bibinfo {year} {2010})}\BibitemShut {NoStop}%
\bibitem [{\citenamefont {Rao}\ and\ \citenamefont
  {Friend}(2017)}]{rao2017harnessing}%
  \BibitemOpen
  \bibfield  {author} {\bibinfo {author} {\bibfnamefont {A.}~\bibnamefont
  {Rao}}\ and\ \bibinfo {author} {\bibfnamefont {R.~H.}\ \bibnamefont
  {Friend}},\ }\href@noop {} {\bibfield  {journal} {\bibinfo  {journal} {Nat.
  Rev. Mater.}\ }\textbf {\bibinfo {volume} {2}},\ \bibinfo {pages} {17063}
  (\bibinfo {year} {2017})}\BibitemShut {NoStop}%
\bibitem [{\citenamefont {Hanna}\ and\ \citenamefont
  {Nozik}(2006)}]{hanna2006solar}%
  \BibitemOpen
  \bibfield  {author} {\bibinfo {author} {\bibfnamefont {M.}~\bibnamefont
  {Hanna}}\ and\ \bibinfo {author} {\bibfnamefont {A.}~\bibnamefont {Nozik}},\
  }\href@noop {} {\bibfield  {journal} {\bibinfo  {journal} {J. Appl. Phys.}\
  }\textbf {\bibinfo {volume} {100}},\ \bibinfo {pages} {074510} (\bibinfo
  {year} {2006})}\BibitemShut {NoStop}%
\bibitem [{\citenamefont {Tayebjee}\ \emph {et~al.}(2012)\citenamefont
  {Tayebjee}, \citenamefont {Gray-Weale},\ and\ \citenamefont
  {Schmidt}}]{tayebjee2012thermodynamic}%
  \BibitemOpen
  \bibfield  {author} {\bibinfo {author} {\bibfnamefont {M.~J.}\ \bibnamefont
  {Tayebjee}}, \bibinfo {author} {\bibfnamefont {A.~A.}\ \bibnamefont
  {Gray-Weale}}, \ and\ \bibinfo {author} {\bibfnamefont {T.~W.}\ \bibnamefont
  {Schmidt}},\ }\href@noop {} {\bibfield  {journal} {\bibinfo  {journal} {J.
  Phys. Chem. Lett.}\ }\textbf {\bibinfo {volume} {3}},\ \bibinfo {pages}
  {2749} (\bibinfo {year} {2012})}\BibitemShut {NoStop}%
\bibitem [{\citenamefont {Einzinger}\ \emph {et~al.}(2019)\citenamefont
  {Einzinger}, \citenamefont {Wu}, \citenamefont {Kompalla}, \citenamefont
  {Smith}, \citenamefont {Perkinson}, \citenamefont {Nienhaus}, \citenamefont
  {Wieghold}, \citenamefont {Congreve}, \citenamefont {Kahn}, \citenamefont
  {Bawendi} \emph {et~al.}}]{einzinger2019sensitization}%
  \BibitemOpen
  \bibfield  {author} {\bibinfo {author} {\bibfnamefont {M.}~\bibnamefont
  {Einzinger}}, \bibinfo {author} {\bibfnamefont {T.}~\bibnamefont {Wu}},
  \bibinfo {author} {\bibfnamefont {J.~F.}\ \bibnamefont {Kompalla}}, \bibinfo
  {author} {\bibfnamefont {H.~L.}\ \bibnamefont {Smith}}, \bibinfo {author}
  {\bibfnamefont {C.~F.}\ \bibnamefont {Perkinson}}, \bibinfo {author}
  {\bibfnamefont {L.}~\bibnamefont {Nienhaus}}, \bibinfo {author}
  {\bibfnamefont {S.}~\bibnamefont {Wieghold}}, \bibinfo {author}
  {\bibfnamefont {D.~N.}\ \bibnamefont {Congreve}}, \bibinfo {author}
  {\bibfnamefont {A.}~\bibnamefont {Kahn}}, \bibinfo {author} {\bibfnamefont
  {M.~G.}\ \bibnamefont {Bawendi}},  \emph {et~al.},\ }\href@noop {} {\bibfield
   {journal} {\bibinfo  {journal} {Nature}\ }\textbf {\bibinfo {volume}
  {571}},\ \bibinfo {pages} {90} (\bibinfo {year} {2019})}\BibitemShut
  {NoStop}%
\bibitem [{\citenamefont {Congreve}\ \emph {et~al.}(2013)\citenamefont
  {Congreve}, \citenamefont {Lee}, \citenamefont {Thompson}, \citenamefont
  {Hontz}, \citenamefont {Yost}, \citenamefont {Reusswig}, \citenamefont
  {Bahlke}, \citenamefont {Reineke}, \citenamefont {Van~Voorhis},\ and\
  \citenamefont {Baldo}}]{congreve2013external}%
  \BibitemOpen
  \bibfield  {author} {\bibinfo {author} {\bibfnamefont {D.~N.}\ \bibnamefont
  {Congreve}}, \bibinfo {author} {\bibfnamefont {J.}~\bibnamefont {Lee}},
  \bibinfo {author} {\bibfnamefont {N.~J.}\ \bibnamefont {Thompson}}, \bibinfo
  {author} {\bibfnamefont {E.}~\bibnamefont {Hontz}}, \bibinfo {author}
  {\bibfnamefont {S.~R.}\ \bibnamefont {Yost}}, \bibinfo {author}
  {\bibfnamefont {P.~D.}\ \bibnamefont {Reusswig}}, \bibinfo {author}
  {\bibfnamefont {M.~E.}\ \bibnamefont {Bahlke}}, \bibinfo {author}
  {\bibfnamefont {S.}~\bibnamefont {Reineke}}, \bibinfo {author} {\bibfnamefont
  {T.}~\bibnamefont {Van~Voorhis}}, \ and\ \bibinfo {author} {\bibfnamefont
  {M.~A.}\ \bibnamefont {Baldo}},\ }\href@noop {} {\bibfield  {journal}
  {\bibinfo  {journal} {Science}\ }\textbf {\bibinfo {volume} {340}},\ \bibinfo
  {pages} {334} (\bibinfo {year} {2013})}\BibitemShut {NoStop}%
\bibitem [{\citenamefont {Cheng}\ \emph {et~al.}(2010)\citenamefont {Cheng},
  \citenamefont {Khoury}, \citenamefont {Clady}, \citenamefont {Tayebjee},
  \citenamefont {Ekins-Daukes}, \citenamefont {Crossley},\ and\ \citenamefont
  {Schmidt}}]{cheng2010efficiency}%
  \BibitemOpen
  \bibfield  {author} {\bibinfo {author} {\bibfnamefont {Y.~Y.}\ \bibnamefont
  {Cheng}}, \bibinfo {author} {\bibfnamefont {T.}~\bibnamefont {Khoury}},
  \bibinfo {author} {\bibfnamefont {R.~G.}\ \bibnamefont {Clady}}, \bibinfo
  {author} {\bibfnamefont {M.~J.}\ \bibnamefont {Tayebjee}}, \bibinfo {author}
  {\bibfnamefont {N.}~\bibnamefont {Ekins-Daukes}}, \bibinfo {author}
  {\bibfnamefont {M.~J.}\ \bibnamefont {Crossley}}, \ and\ \bibinfo {author}
  {\bibfnamefont {T.~W.}\ \bibnamefont {Schmidt}},\ }\href@noop {} {\bibfield
  {journal} {\bibinfo  {journal} {Phys. Chem. Chem. Phys.}\ }\textbf {\bibinfo
  {volume} {12}},\ \bibinfo {pages} {66} (\bibinfo {year} {2010})}\BibitemShut
  {NoStop}%
\bibitem [{\citenamefont {Ravetz}\ \emph {et~al.}(2019)\citenamefont {Ravetz},
  \citenamefont {Pun}, \citenamefont {Churchill}, \citenamefont {Congreve},
  \citenamefont {Rovis},\ and\ \citenamefont {Campos}}]{ravetz2019photoredox}%
  \BibitemOpen
  \bibfield  {author} {\bibinfo {author} {\bibfnamefont {B.~D.}\ \bibnamefont
  {Ravetz}}, \bibinfo {author} {\bibfnamefont {A.~B.}\ \bibnamefont {Pun}},
  \bibinfo {author} {\bibfnamefont {E.~M.}\ \bibnamefont {Churchill}}, \bibinfo
  {author} {\bibfnamefont {D.~N.}\ \bibnamefont {Congreve}}, \bibinfo {author}
  {\bibfnamefont {T.}~\bibnamefont {Rovis}}, \ and\ \bibinfo {author}
  {\bibfnamefont {L.~M.}\ \bibnamefont {Campos}},\ }\href@noop {} {\bibfield
  {journal} {\bibinfo  {journal} {Nature}\ }\textbf {\bibinfo {volume} {565}},\
  \bibinfo {pages} {343} (\bibinfo {year} {2019})}\BibitemShut {NoStop}%
\bibitem [{\citenamefont {Khnayzer}\ \emph {et~al.}(2012)\citenamefont
  {Khnayzer}, \citenamefont {Blumhoff}, \citenamefont {Harrington},
  \citenamefont {Haefele}, \citenamefont {Deng},\ and\ \citenamefont
  {Castellano}}]{khnayzer2012upconversion}%
  \BibitemOpen
  \bibfield  {author} {\bibinfo {author} {\bibfnamefont {R.~S.}\ \bibnamefont
  {Khnayzer}}, \bibinfo {author} {\bibfnamefont {J.}~\bibnamefont {Blumhoff}},
  \bibinfo {author} {\bibfnamefont {J.~A.}\ \bibnamefont {Harrington}},
  \bibinfo {author} {\bibfnamefont {A.}~\bibnamefont {Haefele}}, \bibinfo
  {author} {\bibfnamefont {F.}~\bibnamefont {Deng}}, \ and\ \bibinfo {author}
  {\bibfnamefont {F.~N.}\ \bibnamefont {Castellano}},\ }\href@noop {}
  {\bibfield  {journal} {\bibinfo  {journal} {Chem. Commun.}\ }\textbf
  {\bibinfo {volume} {48}},\ \bibinfo {pages} {209} (\bibinfo {year}
  {2012})}\BibitemShut {NoStop}%
\bibitem [{\citenamefont {Gray}\ \emph {et~al.}(2014)\citenamefont {Gray},
  \citenamefont {Dzebo}, \citenamefont {Abrahamsson}, \citenamefont
  {Albinsson},\ and\ \citenamefont {Moth-Poulsen}}]{gray2014triplet}%
  \BibitemOpen
  \bibfield  {author} {\bibinfo {author} {\bibfnamefont {V.}~\bibnamefont
  {Gray}}, \bibinfo {author} {\bibfnamefont {D.}~\bibnamefont {Dzebo}},
  \bibinfo {author} {\bibfnamefont {M.}~\bibnamefont {Abrahamsson}}, \bibinfo
  {author} {\bibfnamefont {B.}~\bibnamefont {Albinsson}}, \ and\ \bibinfo
  {author} {\bibfnamefont {K.}~\bibnamefont {Moth-Poulsen}},\ }\href@noop {}
  {\bibfield  {journal} {\bibinfo  {journal} {Phys. Chem. Chem. Phys.}\
  }\textbf {\bibinfo {volume} {16}},\ \bibinfo {pages} {10345} (\bibinfo {year}
  {2014})}\BibitemShut {NoStop}%
\bibitem [{\citenamefont {Liu}\ \emph {et~al.}(2012)\citenamefont {Liu},
  \citenamefont {Yang}, \citenamefont {Feng},\ and\ \citenamefont
  {Li}}]{liu2012blue}%
  \BibitemOpen
  \bibfield  {author} {\bibinfo {author} {\bibfnamefont {Q.}~\bibnamefont
  {Liu}}, \bibinfo {author} {\bibfnamefont {T.}~\bibnamefont {Yang}}, \bibinfo
  {author} {\bibfnamefont {W.}~\bibnamefont {Feng}}, \ and\ \bibinfo {author}
  {\bibfnamefont {F.}~\bibnamefont {Li}},\ }\href@noop {} {\bibfield  {journal}
  {\bibinfo  {journal} {J. Am. Chem. Soc.}\ }\textbf {\bibinfo {volume}
  {134}},\ \bibinfo {pages} {5390} (\bibinfo {year} {2012})}\BibitemShut
  {NoStop}%
\bibitem [{\citenamefont {Liu}\ \emph {et~al.}(2018)\citenamefont {Liu},
  \citenamefont {Xu}, \citenamefont {Yang}, \citenamefont {Tian}, \citenamefont
  {Zhang},\ and\ \citenamefont {Li}}]{liu2018highly}%
  \BibitemOpen
  \bibfield  {author} {\bibinfo {author} {\bibfnamefont {Q.}~\bibnamefont
  {Liu}}, \bibinfo {author} {\bibfnamefont {M.}~\bibnamefont {Xu}}, \bibinfo
  {author} {\bibfnamefont {T.}~\bibnamefont {Yang}}, \bibinfo {author}
  {\bibfnamefont {B.}~\bibnamefont {Tian}}, \bibinfo {author} {\bibfnamefont
  {X.}~\bibnamefont {Zhang}}, \ and\ \bibinfo {author} {\bibfnamefont
  {F.}~\bibnamefont {Li}},\ }\href@noop {} {\bibfield  {journal} {\bibinfo
  {journal} {ACS Appl. Mater. Interfaces}\ }\textbf {\bibinfo {volume} {10}},\
  \bibinfo {pages} {9883} (\bibinfo {year} {2018})}\BibitemShut {NoStop}%
\bibitem [{\citenamefont {Tayebjee}\ \emph {et~al.}(2017)\citenamefont
  {Tayebjee}, \citenamefont {Sanders}, \citenamefont {Kumarasamy},
  \citenamefont {Campos}, \citenamefont {Sfeir},\ and\ \citenamefont
  {McCamey}}]{tayebjee2017quintet}%
  \BibitemOpen
  \bibfield  {author} {\bibinfo {author} {\bibfnamefont {M.~J.}\ \bibnamefont
  {Tayebjee}}, \bibinfo {author} {\bibfnamefont {S.~N.}\ \bibnamefont
  {Sanders}}, \bibinfo {author} {\bibfnamefont {E.}~\bibnamefont {Kumarasamy}},
  \bibinfo {author} {\bibfnamefont {L.~M.}\ \bibnamefont {Campos}}, \bibinfo
  {author} {\bibfnamefont {M.~Y.}\ \bibnamefont {Sfeir}}, \ and\ \bibinfo
  {author} {\bibfnamefont {D.~R.}\ \bibnamefont {McCamey}},\ }\href@noop {}
  {\bibfield  {journal} {\bibinfo  {journal} {Nat. Phys.}\ }\textbf {\bibinfo
  {volume} {13}},\ \bibinfo {pages} {182} (\bibinfo {year} {2017})}\BibitemShut
  {NoStop}%
\bibitem [{\citenamefont {Weiss}\ \emph {et~al.}(2017)\citenamefont {Weiss},
  \citenamefont {Bayliss}, \citenamefont {Kraffert}, \citenamefont {Thorley},
  \citenamefont {Anthony}, \citenamefont {Bittl}, \citenamefont {Friend},
  \citenamefont {Rao}, \citenamefont {Greenham},\ and\ \citenamefont
  {Behrends}}]{weiss2017strongly}%
  \BibitemOpen
  \bibfield  {author} {\bibinfo {author} {\bibfnamefont {L.~R.}\ \bibnamefont
  {Weiss}}, \bibinfo {author} {\bibfnamefont {S.~L.}\ \bibnamefont {Bayliss}},
  \bibinfo {author} {\bibfnamefont {F.}~\bibnamefont {Kraffert}}, \bibinfo
  {author} {\bibfnamefont {K.~J.}\ \bibnamefont {Thorley}}, \bibinfo {author}
  {\bibfnamefont {J.~E.}\ \bibnamefont {Anthony}}, \bibinfo {author}
  {\bibfnamefont {R.}~\bibnamefont {Bittl}}, \bibinfo {author} {\bibfnamefont
  {R.~H.}\ \bibnamefont {Friend}}, \bibinfo {author} {\bibfnamefont
  {A.}~\bibnamefont {Rao}}, \bibinfo {author} {\bibfnamefont {N.~C.}\
  \bibnamefont {Greenham}}, \ and\ \bibinfo {author} {\bibfnamefont
  {J.}~\bibnamefont {Behrends}},\ }\href@noop {} {\bibfield  {journal}
  {\bibinfo  {journal} {Nat. Phys.}\ }\textbf {\bibinfo {volume} {13}},\
  \bibinfo {pages} {176} (\bibinfo {year} {2017})}\BibitemShut {NoStop}%
\bibitem [{\citenamefont {Lubert-Perquel}\ \emph {et~al.}(2018)\citenamefont
  {Lubert-Perquel}, \citenamefont {Salvadori}, \citenamefont {Dyson},
  \citenamefont {Stavrinou}, \citenamefont {Montis}, \citenamefont {Nagashima},
  \citenamefont {Kobori}, \citenamefont {Heutz},\ and\ \citenamefont
  {Kay}}]{lubert2018multiple}%
  \BibitemOpen
  \bibfield  {author} {\bibinfo {author} {\bibfnamefont {D.}~\bibnamefont
  {Lubert-Perquel}}, \bibinfo {author} {\bibfnamefont {E.}~\bibnamefont
  {Salvadori}}, \bibinfo {author} {\bibfnamefont {M.}~\bibnamefont {Dyson}},
  \bibinfo {author} {\bibfnamefont {P.~N.}\ \bibnamefont {Stavrinou}}, \bibinfo
  {author} {\bibfnamefont {R.}~\bibnamefont {Montis}}, \bibinfo {author}
  {\bibfnamefont {H.}~\bibnamefont {Nagashima}}, \bibinfo {author}
  {\bibfnamefont {Y.}~\bibnamefont {Kobori}}, \bibinfo {author} {\bibfnamefont
  {S.}~\bibnamefont {Heutz}}, \ and\ \bibinfo {author} {\bibfnamefont {C.~W.}\
  \bibnamefont {Kay}},\ }\href@noop {} {\bibfield  {journal} {\bibinfo
  {journal} {Nat. Commun.}\ }\textbf {\bibinfo {volume} {9}},\ \bibinfo {pages}
  {4222} (\bibinfo {year} {2018})}\BibitemShut {NoStop}%
\bibitem [{\citenamefont {Bayliss}\ \emph {et~al.}(2018)\citenamefont
  {Bayliss}, \citenamefont {Weiss}, \citenamefont {Mitioglu}, \citenamefont
  {Galkowski}, \citenamefont {Yang}, \citenamefont {Yunusova}, \citenamefont
  {Surrente}, \citenamefont {Thorley}, \citenamefont {Behrends}, \citenamefont
  {Bittl} \emph {et~al.}}]{bayliss2018site}%
  \BibitemOpen
  \bibfield  {author} {\bibinfo {author} {\bibfnamefont {S.~L.}\ \bibnamefont
  {Bayliss}}, \bibinfo {author} {\bibfnamefont {L.~R.}\ \bibnamefont {Weiss}},
  \bibinfo {author} {\bibfnamefont {A.}~\bibnamefont {Mitioglu}}, \bibinfo
  {author} {\bibfnamefont {K.}~\bibnamefont {Galkowski}}, \bibinfo {author}
  {\bibfnamefont {Z.}~\bibnamefont {Yang}}, \bibinfo {author} {\bibfnamefont
  {K.}~\bibnamefont {Yunusova}}, \bibinfo {author} {\bibfnamefont
  {A.}~\bibnamefont {Surrente}}, \bibinfo {author} {\bibfnamefont {K.~J.}\
  \bibnamefont {Thorley}}, \bibinfo {author} {\bibfnamefont {J.}~\bibnamefont
  {Behrends}}, \bibinfo {author} {\bibfnamefont {R.}~\bibnamefont {Bittl}},
  \emph {et~al.},\ }\href@noop {} {\bibfield  {journal} {\bibinfo  {journal}
  {Proc. Natl. Acad. Sci. USA}\ }\textbf {\bibinfo {volume} {115}},\ \bibinfo
  {pages} {5077} (\bibinfo {year} {2018})}\BibitemShut {NoStop}%
\bibitem [{\citenamefont {Bayliss}\ \emph {et~al.}(2016)\citenamefont
  {Bayliss}, \citenamefont {Weiss}, \citenamefont {Rao}, \citenamefont
  {Friend}, \citenamefont {Chepelianskii},\ and\ \citenamefont
  {Greenham}}]{bayliss2016spin}%
  \BibitemOpen
  \bibfield  {author} {\bibinfo {author} {\bibfnamefont {S.~L.}\ \bibnamefont
  {Bayliss}}, \bibinfo {author} {\bibfnamefont {L.~R.}\ \bibnamefont {Weiss}},
  \bibinfo {author} {\bibfnamefont {A.}~\bibnamefont {Rao}}, \bibinfo {author}
  {\bibfnamefont {R.~H.}\ \bibnamefont {Friend}}, \bibinfo {author}
  {\bibfnamefont {A.~D.}\ \bibnamefont {Chepelianskii}}, \ and\ \bibinfo
  {author} {\bibfnamefont {N.~C.}\ \bibnamefont {Greenham}},\ }\href@noop {}
  {\bibfield  {journal} {\bibinfo  {journal} {Phys. Rev. B}\ }\textbf {\bibinfo
  {volume} {94}},\ \bibinfo {pages} {045204} (\bibinfo {year}
  {2016})}\BibitemShut {NoStop}%
\bibitem [{\citenamefont {Weil}\ and\ \citenamefont
  {Bolton}(2007)}]{weil2007electron}%
  \BibitemOpen
  \bibfield  {author} {\bibinfo {author} {\bibfnamefont {J.~A.}\ \bibnamefont
  {Weil}}\ and\ \bibinfo {author} {\bibfnamefont {J.~R.}\ \bibnamefont
  {Bolton}},\ }\href@noop {} {\emph {\bibinfo {title} {Electron paramagnetic
  resonance: elementary theory and practical applications}}}\ (\bibinfo
  {publisher} {John Wiley \& Sons},\ \bibinfo {year} {2007})\BibitemShut
  {NoStop}%
\bibitem [{\citenamefont {Benk}\ and\ \citenamefont
  {Sixl}(1981)}]{benk1981theory}%
  \BibitemOpen
  \bibfield  {author} {\bibinfo {author} {\bibfnamefont {H.}~\bibnamefont
  {Benk}}\ and\ \bibinfo {author} {\bibfnamefont {H.}~\bibnamefont {Sixl}},\
  }\href@noop {} {\bibfield  {journal} {\bibinfo  {journal} {Mol. Phys.}\
  }\textbf {\bibinfo {volume} {42}},\ \bibinfo {pages} {779} (\bibinfo {year}
  {1981})}\BibitemShut {NoStop}%
\bibitem [{\citenamefont {Stern}\ \emph {et~al.}(2017)\citenamefont {Stern},
  \citenamefont {Cheminal}, \citenamefont {Yost}, \citenamefont {Broch},
  \citenamefont {Bayliss}, \citenamefont {Chen}, \citenamefont {Tabachnyk},
  \citenamefont {Thorley}, \citenamefont {Greenham}, \citenamefont {Hodgkiss}
  \emph {et~al.}}]{stern2017vibronically}%
  \BibitemOpen
  \bibfield  {author} {\bibinfo {author} {\bibfnamefont {H.~L.}\ \bibnamefont
  {Stern}}, \bibinfo {author} {\bibfnamefont {A.}~\bibnamefont {Cheminal}},
  \bibinfo {author} {\bibfnamefont {S.~R.}\ \bibnamefont {Yost}}, \bibinfo
  {author} {\bibfnamefont {K.}~\bibnamefont {Broch}}, \bibinfo {author}
  {\bibfnamefont {S.~L.}\ \bibnamefont {Bayliss}}, \bibinfo {author}
  {\bibfnamefont {K.}~\bibnamefont {Chen}}, \bibinfo {author} {\bibfnamefont
  {M.}~\bibnamefont {Tabachnyk}}, \bibinfo {author} {\bibfnamefont
  {K.}~\bibnamefont {Thorley}}, \bibinfo {author} {\bibfnamefont
  {N.}~\bibnamefont {Greenham}}, \bibinfo {author} {\bibfnamefont {J.~M.}\
  \bibnamefont {Hodgkiss}},  \emph {et~al.},\ }\href@noop {} {\bibfield
  {journal} {\bibinfo  {journal} {Nat. Chem.}\ }\textbf {\bibinfo {volume}
  {9}},\ \bibinfo {pages} {1205} (\bibinfo {year} {2017})}\BibitemShut
  {NoStop}%
\bibitem [{\citenamefont {Stern}\ \emph {et~al.}(2015)\citenamefont {Stern},
  \citenamefont {Musser}, \citenamefont {Gelinas}, \citenamefont {Parkinson},
  \citenamefont {Herz}, \citenamefont {Bruzek}, \citenamefont {Anthony},
  \citenamefont {Friend},\ and\ \citenamefont
  {Walker}}]{stern2015identification}%
  \BibitemOpen
  \bibfield  {author} {\bibinfo {author} {\bibfnamefont {H.~L.}\ \bibnamefont
  {Stern}}, \bibinfo {author} {\bibfnamefont {A.~J.}\ \bibnamefont {Musser}},
  \bibinfo {author} {\bibfnamefont {S.}~\bibnamefont {Gelinas}}, \bibinfo
  {author} {\bibfnamefont {P.}~\bibnamefont {Parkinson}}, \bibinfo {author}
  {\bibfnamefont {L.~M.}\ \bibnamefont {Herz}}, \bibinfo {author}
  {\bibfnamefont {M.~J.}\ \bibnamefont {Bruzek}}, \bibinfo {author}
  {\bibfnamefont {J.}~\bibnamefont {Anthony}}, \bibinfo {author} {\bibfnamefont
  {R.~H.}\ \bibnamefont {Friend}}, \ and\ \bibinfo {author} {\bibfnamefont
  {B.~J.}\ \bibnamefont {Walker}},\ }\href@noop {} {\bibfield  {journal}
  {\bibinfo  {journal} {Proc. Natl. Acad. Sci. USA}\ }\textbf {\bibinfo
  {volume} {112}},\ \bibinfo {pages} {7656} (\bibinfo {year}
  {2015})}\BibitemShut {NoStop}%
\bibitem [{\citenamefont {Yago}\ \emph {et~al.}(2016)\citenamefont {Yago},
  \citenamefont {Ishikawa}, \citenamefont {Katoh},\ and\ \citenamefont
  {Wakasa}}]{yago2016magnetic}%
  \BibitemOpen
  \bibfield  {author} {\bibinfo {author} {\bibfnamefont {T.}~\bibnamefont
  {Yago}}, \bibinfo {author} {\bibfnamefont {K.}~\bibnamefont {Ishikawa}},
  \bibinfo {author} {\bibfnamefont {R.}~\bibnamefont {Katoh}}, \ and\ \bibinfo
  {author} {\bibfnamefont {M.}~\bibnamefont {Wakasa}},\ }\href@noop {}
  {\bibfield  {journal} {\bibinfo  {journal} {J. Phys. Chem. C}\ }\textbf
  {\bibinfo {volume} {120}},\ \bibinfo {pages} {27858} (\bibinfo {year}
  {2016})}\BibitemShut {NoStop}%
\bibitem [{SM()}]{SM}%
  \BibitemOpen
  \href@noop {} {\ }\bibinfo {note} {See Supplemental Material at [URL will be
  inserted by publisher]}\BibitemShut {NoStop}%
\bibitem [{\citenamefont {Bayliss}\ \emph {et~al.}(2014)\citenamefont
  {Bayliss}, \citenamefont {Chepelianskii}, \citenamefont {Sepe}, \citenamefont
  {Walker}, \citenamefont {Ehrler}, \citenamefont {Bruzek}, \citenamefont
  {Anthony},\ and\ \citenamefont {Greenham}}]{bayliss2014geminate}%
  \BibitemOpen
  \bibfield  {author} {\bibinfo {author} {\bibfnamefont {S.~L.}\ \bibnamefont
  {Bayliss}}, \bibinfo {author} {\bibfnamefont {A.~D.}\ \bibnamefont
  {Chepelianskii}}, \bibinfo {author} {\bibfnamefont {A.}~\bibnamefont {Sepe}},
  \bibinfo {author} {\bibfnamefont {B.~J.}\ \bibnamefont {Walker}}, \bibinfo
  {author} {\bibfnamefont {B.}~\bibnamefont {Ehrler}}, \bibinfo {author}
  {\bibfnamefont {M.~J.}\ \bibnamefont {Bruzek}}, \bibinfo {author}
  {\bibfnamefont {J.~E.}\ \bibnamefont {Anthony}}, \ and\ \bibinfo {author}
  {\bibfnamefont {N.~C.}\ \bibnamefont {Greenham}},\ }\href@noop {} {\bibfield
  {journal} {\bibinfo  {journal} {Phys. Rev. Lett.}\ }\textbf {\bibinfo
  {volume} {112}},\ \bibinfo {pages} {238701} (\bibinfo {year}
  {2014})}\BibitemShut {NoStop}%
\bibitem [{\citenamefont {Eaton}\ \emph {et~al.}(2013)\citenamefont {Eaton},
  \citenamefont {Parkin},\ and\ \citenamefont
  {Anthony}}]{anthony2013Experimental}%
  \BibitemOpen
  \bibfield  {author} {\bibinfo {author} {\bibfnamefont {D.}~\bibnamefont
  {Eaton}}, \bibinfo {author} {\bibfnamefont {S.}~\bibnamefont {Parkin}}, \
  and\ \bibinfo {author} {\bibfnamefont {J.}~\bibnamefont {Anthony}},\ }\href
  {\doibase 10.5517/cc119qsv} {\bibfield  {journal} {\bibinfo  {journal}
  {CCCDC}\ } (\bibinfo {year} {2013}),\ 10.5517/cc119qsv}\BibitemShut {NoStop}%
\end{thebibliography}%

\end{document}